# Quantum phase transitions and superconductivity in the pressurized heavy-fermion compound CeCuP$_2$

Erjian Cheng[1], Chuchu Zhu[1], Tianping Ying[1], Yuanji Xu[2], Darren C. Peets[3], Jiamin Ni[1], Binglin Pan[1], Yeyu Huang[1], Linshu Wang[1], Yi-feng Yang[2,*] and Shiyan Li[1,4,5,*]

[1] *State Key Laboratory of Surface Physics, Department of Physics, Fudan University, Shanghai 200438, China*

[2] *Beijing National Laboratory for Condensed Matter Physics and Institute of Physics, Chinese Academy of Sciences, Beijing 100190, China*

[3] *Ningbo Institute of Materials Technology and Engineering, Chinese Academy of Sciences, Ningbo, Zhejiang 315201, China*

[4] *Collaborative Innovation Center of Advanced Microstructures, Nanjing 210093, China*

[5] *Shanghai Research Center for Quantum Sciences, Shanghai 201315, China*

Corresponding author. Email: yifeng@iphy.ac.cn (Y.-F.Y.); shiyan_li@fudan.edu.cn (S.Y.L.)

**Abstract**

**The tilted balance among competing interactions can yield a rich variety of ground states of quantum matter. In most Ce-based heavy fermion systems, this can often be qualitatively described by the famous Doniach phase diagram, owing to the competition between the Kondo screening and the Ruderman-Kittel-Kasuya-Yoshida exchange interaction. Here, we report an**

**unusual pressure-temperature phase diagram beyond the Doniach one in CeCuP$_2$. At ambient pressure, CeCuP$_2$ displays typical heavy-fermion behavior, albeit with a very low carrier density. With lowering temperature, it shows a crossover from a non-Fermi liquid to a Fermi liquid at around 2.4 K. But surprisingly, the Kondo coherence temperature decreases with increasing pressure, opposite to that in most Ce-based heavy fermion compounds. Upon further compression, two superconducting phases are revealed. At 48.0 GPa, the transition temperature reaches 6.1 K, the highest among all Ce-based heavy fermion superconductors. We argue for possible roles of valence tuning and fluctuations associated with its special crystal structure in addition to the hybridization effect. These unusual phase diagrams suggest that CeCuP$_2$ is a novel platform for studying the rich heavy fermions physics beyond the conventional Doniach paradigm.**

## Introduction

As prototype examples of strongly correlated electron systems, heavy-fermion (HF) compounds have been intensively investigated due to their various fascinating properties [1-4]. They are mostly rare earth or actinide (Ce, Yb, U, etc.) intermetallics with partially filled 4$f$ or 5$f$ orbitals [1-6]. The $f$-electrons are often localized, in particular in Ce-based compounds, and coupled with conduction electrons through a local antiferromagnetic exchange interaction. The resulting Kondo screening effect competes with the induced long-range Ruderman-Kittel-Kasuya-Yoshida (RKKY) interaction between $f$-moments, causing a quantum phase transition between magnetically ordered states and a disordered heavy Fermi liquid state under external tuning such as the pressure, doping or magnetic field [3,7]. This is qualitatively described by the famous Doniach phase diagram [7], as schematically shown in the inset of Fig. 1(b). In most Ce-based HF compounds, the Doniach paradigm has been extensively examined, with only few possible exceptions [3].

In proximity to the magnetic quantum critical point (QCP), superconductivity often emerges, mostly ascribed to the spin-fluctuation-mediated pairing mechanism, and shares some common features with other unconventional superconductors such as the cuprate, iron-based and organic superconductors [3,4]. Other pairing mechanisms such as valence or orbital fluctuations have also been proposed, but no concrete evidences have been established [8-11]. A unified explanation is still lacking and the role of quantum criticality in driving superconductivity is not yet fully understood. Seeking new materials with potentially different low-energy quantum fluctuations and pairing mechanisms may shed light on this long-standing issue.

In this work, we report systematic magnetic susceptibility, specific heat and electrical transport measurements of the heavy-fermion compound $CeCuP_2$. Our results are summarized in the $P$-$T$ phase diagrams of Figs. 1(a) and (b). At ambient pressure, $CeCuP_2$ exhibits typical NFL behavior featuring a quasi-linear temperature-dependent resistivity and divergent specific heat and magnetic susceptibility above 2.4 K. Around 2.4 K, a crossover turns the system into a FL state at low temperatures. Scaling analyses of the magnetic susceptibility and Grüneisen ratio confirm the quantum critical behavior at ambient pressure and zero magnetic field. Under pressure, electrical transport measurements identify multiple quantum phase transitions. For pressure $P < P_{c1} \sim 0.7$ GPa, the system always turns into a metallic Kondo coherent state with lowering temperature. But in contrast to most Ce-based HF compounds, the coherence temperature $T_{coh}$ is gradually suppressed with increasing pressure and diminishes at $P_{c1}$, where the insulating-like Kondo incoherent state persists down to our lowest measured temperature of 0.3 K. For $P_{c1} < P < P_{c2} \sim 4.0$ GPa, the incoherent state is gradually weakened. A second metallic state takes over at low temperatures above $P_{c2}$, accompanied by the first superconducting transition. The $T_c$s for the first superconducting state (SC1) display a clear dome shape with a maximum transition temperature of 4.3 K at 7.8 GPa. At $P_{c3}$ ~ 11.0 GPa, the first superconducting state disappears. With further pressurization beyond $P_{c3}$, the second superconducting state (SC2) arises, and spans over a wide pressure range, which is much broader in pressure than that in usual HF superconductors [3]. The $T_c$ for SC2 exhibits a dip at $P_{c4}$ ~ 34.0 GPa, and then goes up again. At 48.0 GPa, $T_c$ reaches 6.1 K, the highest among all Ce-based heavy fermion superconductors [3]. The overall phase diagram is shown in Figs. 1(a) and (b) and looks strikingly different from the usual Doniach one, which suggests that other essential factors such as the valence change of Ce-ions should be considered for understanding CeCuP$_2$.

## Results

**Non-Fermi liquid behavior and anomalous crossover transition.** The growth of CeCuP$_2$ was first reported by Chykhrij et al. [12] with the space group $I4/mmm$ (139) and the Ce : Cu : P stoichiometric ratio of 1 : 1.09 : 1.8. Each unit cell has one Wyckoff position (4e) for Cerium atoms located between neighboring layers and forming a tetragonal structure. Copper atoms are distributed on two different Wyckoff positions denoted as Cu1 and Cu2, with Cu1 (4d) atoms forming a 12-vertex polyhedron. Phosphorus atoms occupy three Wyckoff positions denoted as P1, P2 and P3. The P1 (4e) and Cu2 (4e) atoms form Platonic solids, and the P2 (4c) and P3 (8j) atoms have distorted cubic co-ordinations. The crystal structure is displayed in the inset of Fig. 1(a). According to the ICSD data (No. 83894) of CeCu$_{1.09}$P$_{1.87}$ built on Chykhrij's work [12], we found fractional occupancies on Cu2, P2, and P3, which explain the non-integral stoichiometric ratio. To determine the accurate chemical composition and crystallographic parameters of our crystals, we have conducted the inductively coupled plasma (ICP) spectroscopy and the Rietveld refinement of powder X-ray diffraction (XRD) studies. Detailed results and discussions can be found in the Supplementary Information (Supplementary Note 1) [13], yielding the Ce : Cu : P ratio of 1 : 1.14 : 2 for our samples. For simplification, we still use $CeCuP_2$ throughout this work.

Single crystals of $CeCuP_2$ were grown from molten Sn flux and demonstrated by XRD measurements (see Supplementary Fig. 1) [13]. Figure 2(a) displays the in-plane resistivity from 0.3 to 4 K at zero magnetic field and shows three distinct regimes. Below ∼1.5 K, the data follows exactly the Fermi liquid power law, $\rho(T) = \rho_0 + AT^n$ with $n = 2$, where $\rho_0$ is the residual resistivity and $A$ the electronic scattering coefficient. Our best fit gives $\rho_0 = 0.058$ mΩ cm and $A = 0.0048$ mΩ cm K$^{-2}$. Above 2.4 K, the resistivity exhibits linear temperature dependence, typical of NFL behavior. In between, there is an abrupt crossover from the NFL to FL states. At higher temperatures as shown in the inset of Fig. 2(a), $CeCuP_2$ behaves like a typical heavy fermion metal, with a resistivity maximum at $T_{coh}$ ∼ 20.5 K marking the transition from a metallic Kondo coherent state to a high-temperature insulating-like Kondo incoherent state. For contrast, the resistivity of $LaCuP_2$ (see Supplementary Fig. 2) is also displayed in the Supplementary Information [13].

To clarify the NFL-to-FL crossover around 2.4 K, we have conducted the specific heat measurements. Figure 2(b) shows the specific heats of $CeCuP_2$, $LaCuP_2$, and the magnetic contributions ($C_m/T$) extracted by comparison against that of the nonmagnetic analogue $LaCuP_2$ from $CeCuP_2$. Herein, we presume that the electron and phonon contributions in $LaCuP_2$ are identical to those in $CeCuP_2$, which is less rigorous due to the special crystal structure. We note that the specific heat in $CeCuP_2$ and $LaCuP_2$ shows similar behavior above the coherence temperature, and in particular both have peaks around $T \sim 100K$. The specific heat of CeCuP$_2$ is larger than that of LaCuP$_2$ over the entire measured range, and this enhancement may stem from the integrated effects of the *f*-electron excitations, the hybridization between the conduction electrons and the excited crystalline electric field (CEF) doublets, and the sample quality. The solid curve in Fig. 2(b) represents the magnetic entropy $S_m$ calculated from $C_m/T$. The effective Kondo temperature ($T_K$), estimated from $S_m(T_K) = 0.65\ R\ln2$ [14], is about 9.5 K. $S_m$ reaches 0.38 $R\ln2$ at the low-temperature hump and the remaining entropy of the Kramers doublet is gradually released till 29 K somewhat above $T_{coh}$. The green points in Fig. 2(b) shows the resulting $C_m/T$ of CeCuP$_2$ at zero field, which is seen to diverge logarithmically below 10 K but interrupted by a hump at $\sim 1.8$ K. The hump causes a departure from the logarithmic divergence below 2.4 K (marked as $T^*$). This is in exact correspondence with the NFL-to-FL crossover in the resistivity. The hump cannot be described by a two-level Schottky function (Supplementary Fig. 3(a)) [13], excluding the possibility of a Schottky anomaly indirectly. The shape of the hump also does not follow a distinct first-order '$\lambda$'-anomaly or a jump characteristic of a second-order phase transition; therefore, it likely cannot be attributed to a long-range magnetic order. A similar feature has also been found in the arsenic analogue CeCuAs$_2$ around 4 K, where it has been explained as a disorder-driven continuous metal-insulator transition resulting in a gradual loss of density of states (DOS) with decreasing temperature, although the low-temperature downturn in $C/T$ in CeCuAs$_2$ is still not well understood [15,16].

The absence of long-range magnetic order is more clearly demonstrated in Fig. 2(c), where the temperature dependence of the magnetic susceptibility is plotted down to 1.8 K, and below 2 K (left inset) measured in a $^3$He cryostat. Short-range magnetic order is also excluded by field cooling (FC) and zero FC (ZFC) (Fig. S4(a)), and *ac* magnetic susceptibility measurements (Supplementary Fig.

4(b)) [13]. In CeCuAs$_2$, spin-glass freezing and Schottky anomaly origin behind the low-temperature hump in specific heat have also been excluded [14], as we did for CeCuP$_2$. The right inset gives the inverse susceptibility at 0.5 T. By fitting the linear part of the data with the Curie-Weiss law, $\chi = C/(T-\theta)$, we obtain $\theta \sim$ -284 K and an effective magnetic moment ($\mu_{eff}$) of 1.44 $\mu_B$. The former indicates an effective antiferromagnetic exchange interaction between $f$-moments, but the size of the moment is significantly reduced compared to the expected value of 2.54 $\mu_B$ for a free Ce$^{3+}$ ion, suggesting possible mixed-valence of the Ce ions in CeCuP$_2$. Interestingly, the magnetic susceptibility at 0.05 T exhibits an evident power-law divergence with $T^{-0.71}$ at low temperatures. A comparable exponent has been observed previously in YbRh$_2$(Si$_{0.95}$Ge$_{0.05}$)$_2$ with $T^{-(0.6 \pm 0.1)}$ and explained in terms of local quantum criticality [17]. The similarity suggests potential magnetic quantum-critical fluctuations in CeCuP$_2$ at ambient pressure.

More insight on the NFL-to-FL crossover can be obtained from the Hall measurements (Supplementary Fig. 5) [13]. Figure 2(d) shows the temperature-dependent Hall coefficient ($R_H$) obtained by fitting the Hall resistivity from 0 to 3 T. We found dominant electron carriers from 0.3 to 20 K. With lowering temperature to 4 K, the magnitude of $R_H$ increases monotonously, indicating the increasement of concentration of electron carriers. Upon decreasing the temperature to $\sim$ 2.4 K, a peak in $R_H$ emerges, and then $R_H$ decreases. The peak observed in $R_H$ may arise from phase transitions smeared by disorders. Since magnetic ordering has been excluded in our experiments, the peak in $R_H$ possibly comes from structural variations in consideration of the fractional occupancies on Cu2, P2 and P3 [18,19]. At the lowest measured temperature, the density of electrons can be estimated to be about 0.006 $e^-$ per formula unit, ten times smaller than that in usual HF metals such as CeCoIn$_5$. According to Nozières' argument [20,21], this may produce a protracted Kondo screening [22,23], as the *f*-moments cannot be fully screened. In CeCuP$_2$, we anticipate that the special electronic structures of the fractionally occupied Cu2, P2, and P3 atoms may tune the Fermi level and leave only few carriers in the conduction bands. Thus, CeCuP$_2$ provides a unique platform for valence tuning in HF materials. The NFL-to-FL anomalies may be associated with this special property of CeCuP$_2$, an interesting topic to be investigated using the state-of-the-art angle-resolved photoemission spectroscopy (ARPES) or scanning tunneling microscopy (STM) in the future.

The above anomalous features are also reflected in the field measurements. Figure 3(a) plots the in-plane resistivity as a function of the magnetic field applied along the *c* axis. All resistivity curves cross roughly at a single point at ∼ 2.4 K, around which the magnetoresistance (MR) changes sign. This coincides with the abrupt change in the carrier concentration. The crossing point shifts slightly to higher temperature with increasing field. Isothermal MR measurements (Supplementary Fig. 6(a)) demonstrate the evolution of the MR [13]. In Fig. 3(b), the hump in the field-dependent specific heat also shifts gradually to higher temperatures, as observed in CeCuAs$_2$ [15]. The overall suppression of the specific heat indicates the suppression of quasi-particle excitations with increasing field [13].

The field measurements provide crucial information on the QCP in CeCuP$_2$ at ambient pressure. To investigate the origin of the NFL above 2.4 K, we measured the magnetic susceptibility under high fields (see Supplementary Fig. 4(c)). In many HF systems, the critical scaling may be described within a fermion-condensed picture [24-26]. Motivated by the established scaling method [25,26], we identify a scaling of its free energy in the scaling variable $x = T^*(B) \sim B^{0.4}$. The exponent of *B* is acquired by fitting the maximum value of $-d\chi/dT$ in different fields. Its value of 0.4 is smaller than in other QC systems, for example, 1 in *β*-YbAlB$_4$ [23], 0.59 in YFe$_2$Al$_{10}$ [26], and 1.05 in CeNi$_2$Ge$_2$ [27]. As shown in Fig. 3(c), the magnetic susceptibility in different fields shows a remarkable scaling collapse for 1.8 K ≤ T ≤ 18 K and B ≤ 7 T, extending over about three orders of magnitude in the scaling variable x, where $\frac{-d\chi}{dT}B^{0.71} = \Psi\left(\frac{T}{B^{0.4}}\right)$. This scaling behavior reveals an important property of the underlying free energy. Inspired by the scaling functions $f_F(x)$ that were found in β-YbAlB$_4$ [25] and YFe$_2$Al$_{10}$ [26], we propose a similar expression for CeCuP$_2$— $M = B^{1-\varepsilon+y}f_M(x)$ with $f_M(x) = c(x^\alpha + a^2)^{\frac{-\varepsilon}{2}}$ and $x = T/B^y$. The scaling function $f_M$ is related to the original scaling function of the free energy $f_F$, as described in Ref. 26, and y is a magnetic field tuning parameter. By taking the temperature derivative of $\chi = M/B$, we deduce the scaling equation—$\frac{-d\chi}{dT}B^\varepsilon = \frac{ca\varepsilon}{2}x^{\alpha-1}(x^\alpha+a^2)^{\frac{-\varepsilon}{2}-1}$. Here, c, a, α and ε are the fitting parameters. In the limit of a field-driven FL phase (x << 1), the parameter α is 2 [25,26]. As shown in Fig. 3(c), the scaling equation can fit the data very well in the limit of quantum criticality (x >>1) with α = 2, but fails in the limit of the field-driven FL phase it would be expected to describe. The reason for this failure may be ascribed to which the system is in the mixed-valence state under ambient pressure. The parameters c, a and ε from the fit with α = 2 are 0.026 ± 0.001, 1.842 ± 0.005, and 0.75 ± 0.05, respectively. The parameter ε = 0.71 can be determined experimentally (see Fig. 2(c)), thus the data agrees well with each other within the fitting uncertainty. We can also treat α as a free parameter to fit the data for all x, yielding an excellent fit. The parameters of c, a, α and ε in this case are 0.024 ±0.001, 2.743 ±0.003, 2.80 ±0.01, and 0.6 ± 0.1, respectively, and ε is again consistent with the experimental value. In contrast to the exponent of x, the parameter α = 2.8 is significantly larger than that expected for a FL phase. To explain this, more theoretical efforts will be required.

A $T^{-2.00(2)}$ divergence is also found in the Grüneisen ratio divided by the magnetic field, as shown in the inset to Fig. 3(c). This observation implies further critical behavior [25,26], although the validity of establishing the QCPs through the Grüneisen ratio is still under debate [28]. Putting together, all evidence supports quantum criticality possibly interrupted by the NFL-to-FL crossover at zero field and ambient pressure in CeCuP$_2$. We have also checked the field dependence of the resistivity coefficient $A$ and the upper boundary of the Fermi liquid regime $T_0$. As shown in Fig. 3(d), $T_0$ changes slightly from 1.7 to 1.5 and $A$ also decreases, suggesting reduced electronic scatterings.

**Pressure-induced quantum phase transitions and superconductivity.** Next, we turn to pressure measurements. Due to the low magnitude of the energy scales in HF materials, pressure is an efficient tool to tune the ground state properties. Figures 4(a) and 4(b) show the resistivity under pressure up to 2.16 GPa measured from 2 to 300 K in PPMS and from 0.3 to 5 K using a $^3$He refrigerator, respectively. With increasing pressure up to $P_{c1}$ ~ 0.7 GPa, $T_{coh}$ is gradually suppressed to zero and the magnitude of the resistivity below $T_{coh}$ increases correspondingly. The resistivity exhibits semi-metallic behavior at 0.71 GPa (left panel of Fig. 4(b)) and its value at 0.3 K is about 22 times larger than that at ambient pressure. The inset of Fig. S7 summarizes the evolution of $T_{coh}$ and the resistivity at 0.3 K with increasing pressure [13]. It is well-known that $T_{coh}$ marks the emergence of heavy quasiparticles and its magnitude reflects the effective strength of collective hybridization between $f$ and conduction electrons [29,30]. The decrease of $T_{coh}$ in CeCuP$_2$ is in sharp contrast to most Ce-based HF materials, where pressure usually enhances the hybridization. In the Doniach paradigm, this is caused by the increase of the Kondo coupling $|J|$, namely $\partial|J|/\partial P > 0$ [30,31]. The situation in CeCuP$_2$ is therefore highly unusual. Similar suppression has also been observed in CeRhIn$_5$ below 1.2 GPa [31], but ascribed to other effects such as the spin freezing or enhanced 2D AFM fluctuations associated with its low-temperature AFM order [31], and UTe$_2$ [32], the increase in valence towards a U$^{4+}$ (5f$^2$) state, which is generally expected to drive the system from superconducting to antiferromagnetic [32]. In Yb-based HF compounds, $T_{coh}$ also decreases with pressure due to the shift of the $f$ levels away from the Fermi energy, which, would typically lead to magnetic ordering at high pressure. Regardless of Ce-based, U-based or Yb-based, the reduced coherence temperature is proposed to be related to magnetic fluctuations. Therefore, the decrease in the Kondo coherence temperature in CeCuP$_2$ motivates us to check the magnetic field effect under pressure. Supplementary Figure 6(b) shows the pressure dependence of the MR investigated at 2 K (detail discussions can be found in the Supplementary Note 7 [13]), finding the change of the scattering mechanism, indicative of the possible magnetic fluctuations, although there is no any indication for magnetic transitions in resistivity under pressure (Figs. 4(a) and 4(b)). In Supplementary Fig. 10, we demonstrate the existence of pressure-induced valence fluctuations of Ce 4f electrons in CeCuP$_2$ by means of the scaling of resistivity under pressure. As a result, the decrease of coherence temperature and the change of MR under pressure are maybe related to the pressure-induced fluctuations of the Ce 4f electrons in CeCuP$_2$. To confirm this proposal, more works are expected to investigate the magnetic behavior under pressure in the future.

The critical behavior around $P_{c1}$ is also different. For a pressure-induced magnetic QCP, the electronic scattering coefficient $A$ usually diverges in the vicinity of the QCP [3], but this is not observed in CeCuP$_2$. Rather, we found a moderate enhancement near $P_{c1}$ (Supplementary Fig. 7) [13]. This suggests that the coherent-to-incoherent transition at $P_{c1}$ is not associated with a usual magnetic or delocalization transition often observed in other HF compounds. Upon further increasing pressure up to $P_{c2} \sim 4.0$ GPa, the resistivity becomes suppressed but the system remains in an incoherent state down to the lowest measured temperature. In this intermediate region, the resistivity responds very differently to the magnetic field compared to the low-pressure region. Fig. 4(c) shows the resistivity at 2.16 GPa, which decreases monotonically with increasing field, indicating a change in the scattering mechanism. Thus, the phase diagram of $CeCuP_2$ cannot be simply understood within the Doniach paradigm based on the competition between long-range magnetism and Kondo hybridization. It is essential to take other important factors into consideration, for example, charge fluctuations.

HF superconductivity has often been observed in the vicinity of magnetic or valance instabilities [3]. Inspired by this, we have carried out further pressure measurements on $CeCuP_2$ polycrystal obtained by crushing a single crystal. Figure 4(d) plots the temperature dependence of the normalized resistance at different pressures without pressure transmitting medium (denoted as Run1). The Kondo incoherent phase persists up to 4.6 GPa, above which a small drop in the resistance is seen to emerge at low temperatures. At 6.3 GPa, the resistance starts to display typical metallic behavior. Surprisingly, a superconducting drop emerges around the same pressure. The superconducting temperature ($T_c^{onset}$) defined by the onset of the transition reaches 4.7 K at 14.8 GPa, as shown in the inset. To verify the superconductivity observed in $CeCuP_2$, pressure experiments were repeated in another device with NaCl as the pressure transmitting pressure (denoted as Run2), as displayed in Fig. 4(e). A small drop in the resistance emerges at 4.0 GPa. With increasing pressure to 7.8 GPa, $T_c$ increases to a maximum of 4.3 K, and then decreases beyond 7.8 GPa. At 11.0 GPa and down to 1.9 K the lowest temperature we measured, there is no any indications of superconductivity. The pressure-dependent $T_c$ ranging from 4.0 to 11.0 GPa displays a clear dome shape, as plotted in Fig. 1(b).

Upon further compression beyond 11.0 GPa, the second superconducting state (SC2) arises, it spans over a wide pressure range, which is much broader in pressure than that in usual HF superconductors [3]. At ~ 34.0 GPa, the $T_c$ displays a dip, which is a mystery for now. At 48.0 GPa the highest pressure we measured, $T_c$ reaches 6.1 K [3]. To verify the superconducting transition and obtain the upper critical fields ($\mu_0H_{c2}$s), we applied the magnetic field at several pressures. As shown in Supplementary Fig. 9, the transition is gradually suppressed by magnetic field, as expected for superconductivity, although zero resistance hasn't been achieved. Considering the fractional occupancies of the crystal sites in CeCuP$_2$, we propose several scenarios for the nonzero resistance. The first one is the poor grain contact in the polycrystalline sample and an anisotropic superconducting order parameter [33]. A single crystal high-pressure transport study may thus help to find zero resistance, and an even higher transition temperature in the superconducting phases of CeCuP$_2$. The second is that the actual stoichiometric ratio of Ce : Cu : P = 1 : 1.14 : 2 is not optimal for superconductivity. The crystal structure of CeCuP$_2$ provides a unique platform for chemical doping. Therefore, tuning the concentration of copper and phosphorus may facilitate the observation of zero resistance. The third is the local inhomogeneity under pressure, which would be associated with a spatial variation of the copper and phosphorus concentration, local density of states, and superconducting order parameter [34]. This inhomogeneity would lead to filamentary superconductivity [34]. The last one is that the pressure is inhomogeneous and not locally hydrostatic. This could drastically smear out the transition, especially if the order parameter is anisotropic and especially at higher pressures. Determining the reason for nonzero resistance is left for a future study. Previously, the highest transition temperatures reported among all Ce-based HF superconductors are 2.3 K for CeCoIn$_5$ at ambient pressure [35] and 2.5 K for CeAu$_2$Si$_2$ at ~ 22.5 GPa [9]. Our observed transition temperature in CeCuP$_2$ is the new highest record among all Ce-based HF superconductors.

To understand the pairing mechanism, we plot in Fig. 4(f) the temperature dependence of $\mu_0H_{c2}$

and use the Ginzburg-Landau (GL) formula $\mu_0H_{c2}(T) = \mu_0H_{c2}(0)(1-(T/T_c)^2)/(1+(T/T_c)^2)$ to fit the data. The values are estimated to be 0.050(6) T, 0.0192(9) T, 0.40(2) T, 0.37(2) T, 1.85(5) T and 2.07(7) T for 8.4 GPa, 9.7 GPa, 15.4 GPa, $T_c^{onset}$ at 23.3 GPa, $T_c^{10\%}$ at 23.3 GPa (denoted as 23.3´GPa) and 42.5 GPa, respectively, which are much lower than the Pauli limiting fields $H_P(0) = 1.84T_c \sim$ 7.9, 6.9, 7.7, 8.0, 5.9 and 10.3 T, respectively, indicating that Pauli pair breaking is not relevant [36,37]. To get some insight into the superconducting pairing, the initial slopes of the upper critical field at 8.4, 9.7, 15.4, 23.3 and 42.5 GPa have been calculated, yielding -0.012, -0.005, -0.096, -0.428 and -0.372 T/K, respectively. The magnitudes of the initial slopes are much smaller than in other heavy-fermion superconductors [3], suggestive of conventional superconducting pairing. However, this quantity can be affected by other factors, such as disorder, so a small initial slope of the upper critical field cannot serve as a smoking gun for the conventional superconducting pairing. For HF superconductors, a $\mu_0H_{c2}$ not exceeding $H_P(0)$ can be also found in other Ce-based systems, for example, $CeRh_2Si_2$ [38], $CePd_2Si_2$ [39], $Ce_3PtIn_{11}$ [40]. To better show the $\mu_0H_{c2}$ behavior of $CeCuP_2$ among Ce-based HF superconductors, we list the ratio of $\mu_0H_{c2}(0)$ to $H_P(0)$, as displayed in Supplementary Fig. 9. For $CeRh_2Si_2$, the resistivity decreases below 0.5 K at ~ 1.05 GPa, and the $\mu_0H_{c2}$ is estimated to be 0.28 T, less than 0.92 T calculated from the Pauli rule [38]. Therefore, the low $\mu_0H_{c2}$s in $CeCuP_2$ could not exclude the possibility of unconventional superconducting pairing completely, although the possibility of BCS-like origin cannot be ruled out either. We also note that the low value of upper critical field in $CeCuP_2$ may suggest flux pinning at grain boundaries. In spite of this, two separated superconducting phases observed in $CeCuP_2$ are quite unconventional [3,41,42].

Often, superconductivity near an AFM QCP as in $CeCoIn_5$ may be ascribed to a spin-fluctuation-mediated pairing mechanism [3]. In $CeCuP_2$, however, the superconductivity develops on the far side of the phase diagram. At ambient pressure, there is no evidence of superconductivity down to the lowest measured temperature. If and to what extent the $f$ electrons may participate in the Cooper pairing remains to be examined in the future. On the other hand, valence fluctuations may also play a role and even promote the pairing [3,41]. In $CeCu_2(Si_{1-x}Ge_x)_2$ and $CeAu_2Si_2$, superconductivity extends to very high pressure far away from the magnetic QCP with a greatly enhanced $T_c$ [3,8,9]. It is thus speculated that superconductivity in $CeCuP_2$ near $P_{c2}$ or $P_{c3}$ may potentially involve a valence mechanism.

**Discussion**

Putting together, our systematic measurements can be summarized in the overall $P$-$T$ phase diagram presented in Figs. 1(a) and 1(b), indicating a rich variation in the electronic properties of $CeCuP_2$ at ambient and high pressure, which is not only different from most other Ce-based HF compounds but also beyond the conventional Doniach paradigm. At ambient pressure, a crossover drives this system from NFL to FL, which is not well understood so far. Scaling analysis demonstrates the quantum critical behavior at ambient pressure and zero field. With increasing pressure, the Kondo coherence state is first suppressed, in contrast to most Ce-based HF compounds, and an intermediate Kondo semi-metallic state emerges in the intermediate pressure region, before a second metallic state emerges at even higher pressures. The unusual phase diagram may be associated with the particular crystal, low-carrier and mixed-valence properties of $CeCuP_2$. The fractional occupancies of the crystal sites provide a rare platform apt for chemical tuning. On the other hand, the low carrier density usually favors a magnetically ordered state. Recent experiments on the HF material $CeNi_{2-\delta}As_2$ with a low carrier density of $\sim$ 0.032 $e^-$ per formula unit observed a pressure-induced quantum phase transition from an AFM order to a coherent Kondo screening state [22], with $T_{coh}$ emerging and increasing with pressure across the QCP, in line with predictions of the Doniach picture. Thus, the low carrier density alone cannot explain the unusual phase diagram in CeCuP$_2$ and it might be essential to consider other key factors.

Since the Ce ions are in a mixed-valence state, we speculate that not only the Kondo hybridization but also the Ce valence is tuned by pressure, which would in turn affect the carrier concentration and the ground state of CeCuP$_2$. As a matter of fact, following previous analysis on CeCu$_2$Si$_2$ and CeRhGe$_3$ [43,44], we find a rough resistivity scaling that tentatively points towards a potential pressure-induced valence transition near $P_{c1}$ in CeCuP$_2$ (Supplementary Fig. 10). More details can be found in the Supplementary Information [13]. If this is the case, the system can no longer be described by the usual Kondo lattice model, which ignores the charge degrees of freedom of the Ce-ions [3]. This might cause the breakdown of the Doniach picture. As proposed by Hattori, pressure can drive the ground state of HFs from heavy Fermi liquid to "light" Fermi liquid, where the effective mass is smaller than that in the former, across a first-order meta-orbital transition based on both a two-orbital Anderson lattice model and the self-consistent renormalization (SCR) theory [11]. For CeCuP$_2$, the temperature dependence of the resistivity displays typical heavy-fermion behavior. Under high pressure where the superconducting transitions occur, the resistivity shows typical metallic behavior, and the *f*-electrons may be partially decoupled from the conduction electrons and localized. The small magnitude of the initial upper critical field slopes further indicates conventional superconducting pairing as well as a "light" effective mass compared with other HF superconductors, although this quantity can be affected by disorder and the nonzero resistance. We nonetheless get strong hints that CeCuP$_2$ is on the left side of the first-order transition line, and pressure drives it from heavy Fermi liquid to "light" Fermi liquid across the transition. In the temperature dependence of the resistivity, the valence fluctuations lead to a $T^n$ (n = 1, 4/3, 5/3) dependence near the critical end point (CEP). We then check the temperature dependence of the resistivity ($R/R_{300K}$) for CeCuP$_2$ at several critical pressures, as shown in Supplementary Fig. 11, finding a $T^n$ (n > 2) dependence, which is beyond the SCR picture. Moreover, the *P-T* phase diagrams obtained here is also different from other Ce-based HF compounds with potential valence instabilities, e.g., CeCu$_2$(Si$_{1-x}$Ge$_x$)$_2$, CeAu$_2$Si$_2$, and possibly also CeIrIn$_5$ [3,8,9,41]. It therefore remains to be understood with more sophisticated models. The superconductivity with enhanced $T_c$ at high pressures might only be fully explained when the valence or other effects are taken into consideration. All these special features make CeCuP$_2$ a unique platform for studying the rich variety of heavy fermion physics beyond the usual Doniach paradigm.

## Methods

**Sample preparation.** High-purity starting materials of Ce powder (99.9%, Alfa Aesar), Cu powder (99.9%, Alfa Aesar), P flakes (99.999 %, Aladdin) and Sn powder (99.99%, Aladdin) were mixed in a molar ratio of 1 : 1 : 10 : 60 and loaded into an alumina crucible, which was sealed under vacuum in a quartz ampoule. The sealed ampoule was very slowly heated to 1150°C and kept there for 3 days, then cooled at a rate of 1.5 °C/h. The ampoule was taken out and the excess Sn flux decanted with a centrifuge at 500°C. LaCuP$_2$ is prepared by an arc-melting method. High-purity starting materials of La chunk (99.9%, Alfa Aesar), Cu lump (99.9%, Alfa Aesar), P flakes (99.999%, Aladdin) in a molar ratio of 1 : 1.15 : 5 were loaded in a water-cooled copper boat, and the mixture was then arc-melted several times. The obtained compound was ground in a glove box, and the powder was pressed into pills. These pills were sealed in a quartz ampoule, and then annealed at 1000°C for 3 days. Chemical composition and crystal structure determinations can be found in the Supplementary Information [13].

**Electrical transport and thermodynamics measurements.** Resistance measurements from 300 to 2 K and below 2 K were performed in a physical property measurement system (PPMS; Quantum Design) and a $^3$He cryostat, respectively. *dc* magnetic susceptibility measurements were performed in a magnetic property measurement system (MPMS; Quantum Design) equipped with a $^3$He cryostat, with magnetic field applied perpendicular to the *ab* plane of the CeCuP$_2$ single crystal. *ac* magnetic susceptibility measurements were performed in a PPMS equipped with a dilution refrigerator. Heat capacity measurements were conducted using the relaxation method in a PPMS equipped with a $^3$He cryostat and magnetic field was also applied along *c* axis of the CeCuP$_2$ single crystal.

**Pressure measurements.** For high-pressure experiments, samples were loaded in a piston-cylinder clamp cell made of Be-Cu alloy, with Daphne oil as the pressure medium. The pressure inside the cell was determined from the $T_c$ of a tin wire. A CeCuP$_2$ single crystal was cut into a bar shape, and the standard four-probe method was used for resistivity measurements, with contacts made using silver epoxy. Higher-pressure measurements were performed on powder samples comprising crushed single crystals using a diamond anvil cell (DAC). For Run1 measurements, no pressure transmitting medium was adopted, while, for Run2 measurements, NaCl was used as the pressure transmitting medium. The experimental pressures were determined by the pressure-induced fluorescence shift of ruby [2] at room temperature before and after each experiment. A direct-current van der Pauw technique was adopted.

## Data availability

The data that support the findings of this study are available from the corresponding author upon reasonable request.

## References

1. Hewson, A. C. & Coleman, P. The Kondo Problem of Heavy Fermions. *Physics Today* **47**, 60-61(1993).

2. Si, Q. M. & Steglich, F. Heavy fermions and quantum phase transitions. *Science* **329**, 1161-1166 (2010).

3. Weng, Z. F. *et al*. Multiple quantum phase transitions and superconductivity in Ce-based heavy fermions. *Rep. Prog. Phys.* **79**, 094503 (2016).

4. Phan, V.-N. Superconductivity in Ce-based heavy-fermion systems under high pressure. *Phys. Rev. B* **101**, 245120 (2020).

5. Helm, T. Non-monotonic pressure dependence of high-field nematicity and magnetism in $CeRhIn_5$. *Nat. Commun.* 11, 3482 (2020).

6. Braithwaite, D. Multiple superconducting phases in a nearly ferromagnetic system. *Commun. Phys.* **2**, 147 (2019).

7. Doniach, S. The Kondo lattice and weak antiferromagnetism. *Physica B+ C* **91**, 231-234 (1977).

8. Yuan, H. Q. *et al*. Non-Fermi liquid states in the pressurized $CeCu_2(Si_{1-x}Ge_x)_2$ system: two critical points. *Phys. Rev. Lett.* **96**, 047008 (2006).

9. Ren, Z. *et al*. Giant overlap between the magnetic and superconducting phases of CeAu$_2$Si$_2$ under pressure. *Phys. Rev. X* **4**, 031055 (2014).

10. Pourovskii, L. V., Hansmann, P., Ferrero, M. & Georges, A. Theoretical prediction and spectroscopic fingerprints of an orbital transition in CeCu$_2$Si$_2$. *Phys. Rev. Lett.* **112**, 106407 (2014).

11. Hattori, K. Meta-orbital transition in heavy-fermion systems: analysis by dynamical mean field theory and self-consistent renormalization theory of orbital fluctuations. *J. Phys. Soc. Jpn.* **79**, 114717 (2010).

12. Chykhrij, S. I., Loukashouk, G. V., Oryshchyn, S. V., Kuz'ma, Y. B. Phase equilibria and crystal structure of compounds in the Ce-Cu-P system. *J. Alloy. Compd.* **248**, 224-232 (1997).

13. Supplementary Information to 'Quantum phase transitions and superconductivity in the pressurized heavy-fermion compound CeCuP$_2$'.

14. Desgranges, H. & Schotte, K. Specific heat of the Kondo model. *Phys. Lett. A* **91**, 240-242 (1982).

15. Sengupta, K. *et al*. Magnetic, electrical resistivity, heat-capacity, and thermopower anomalies in CeCuAs$_2$. *Phys. Rev. B* **70**, 064406 (2004).

16. Cvitkovich, L. *et al*. Anisotropic Physical Properties of the Kondo Semimetal CeCu$_{1.11}$As$_2$. *J. Phys. Soc. Jpn.* **30**. 011020 (2020).

17. Gegenwart, P. *et al*. Ferromagnetic quantum critical fluctuations in YbRh$_2$(Si$_{0.95}$Ge$_{0.05}$)$_2$. *Phys. Rev. Lett.* **94**, 076402 (2005).

18. Ernsting, D. *et al*. Vacancies, disorder-induced smearing of the electronic structure, and its implications for the superconductivity of anti-perovskite $MgC_{0.93}Ni_{2.85}$. *Sci. Rep.* **7**, 10148 (2017).

19. Kuo, C. N. *et al*. Lattice distortion associated with Fermi-surface reconstruction in $Sr_3Rh_4Sn$. *Phys. Rev. B* **91**, 165141 (2015).

20. Nozières, P. Magnetic impurities and the Kondo effect. *Ann. Phys. Fr.* **10**, 19-35 (1985).

21. Nozières, P. Some comments on Kondo lattices and the Mott transition. *Eur. Phys. J. B* **6**, 447-457 (1998).

22. Luo, Y. K. *et al*. Pressure-tuned quantum criticality in the antiferromagnetic Kondo semimetal $CeNi_{2-\delta}As_2$. *P. Natl. Acad. Sci. USA* **112**, 13520-13524 (2015).

23. Zhang, P. *et al*. Protracted Kondo coherence with dilute carrier density in Cerium based nickel pnictides. Preprint at: http://arxiv.org/abs/1909.13251 (2019).

24. Shaginyan, V. R., Amusia, M. Y., Msezane, A. Z., Popov, K. G. Scaling behaviour of heavy fermion metals. *Phys. Rep.* **492**, 31-109 (2010).

25. Matsumoto, Y. *et al*. Quantum criticality without tuning in the mixed valence compound *β*-$YbAlB_4$. *Science* **331**, 316-319 (2011).

26. Wu, L. S. *et al*. Quantum critical fluctuations in layered $YFe_2Al_{10}$. *P. Natl. Acad. Sci. USA* **111**, 14088-14093 (2014).

27. Koerner, S. *et al*. Crossover to Fermi-liquid behavior at lowest temperatures in pure $CeNi_2Ge_2$. *J. Low Temp. Phys.* **119**, 147-153 (2000).

28. Gegenwart, P. Grüneisen parameter studies on heavy fermion quantum criticality. *Rep. Prog. Phys.* **79**, 114502 (2016).

29. Yang, Y.-F. *et al*. Scaling the Kondo lattice. *Nature* **454**, 611-613 (2008).

30. Yang, Y.-F., Pines, D. & Lonzarich, G. Quantum critical scaling and fluctuations in Kondo lattice materials. *P. Natl. Acad. Sci. USA* **114**, 6250-6255 (2017).

31. Hegger, H. *et al*. Pressure-induced superconductivity in quasi-2D CeRhIn$_5$. *Phys. Rev. Lett.* **84(21)**, 4986 (2000).

32. Thomas, S. M. *et al.* Evidence for a pressure-induced antiferromagnetic quantum critical point in intermediate valence UTe$_2$. *Sci. Adv.* **6**, eabc8709 (2020).

33. Mani, A. *et al.* Pressure-induced superconductivity in BaFe$_2$As$_2$ single crystal. *EPL* **87**, 17004 (2009).

34. Gofryk, K. *et al.* Local inhomogeneity and filamentary superconductivity in Pr-doped CaFe$_2$As$_2$. *Phys. Rev. Lett.* **112**, 047005 (2014).

35. Petrovic, C. *et al*. Heavy-fermion superconductivity in CeCoIn$_5$ at 2.3 K. *J. Phys.: Condens. Matter* **13**, L337-L342 (2001).

36. Clogston, A. M. Upper limit for the critical field in hard superconductors. *Phys. Rev. Lett.* **9**, 266-267 (1962).

37. Chandrasekhar, B. S. A note on the maximum critical field of high-field superconductors. *Appl. Phys. Lett.* **1**, 7-8 (1962).

38. Pressure-induced superconductivity in an antiferromagnet CeRh$_2$Si$_2$. *J. Phys.: Condens. Matter* **14**, L377–L383 (2002).

39. Sheikin, I. *et al*. Superconductivity, upper critical field and anomalous normal state in CePd$_2$Si$_2$ near the quantum critical point. *J. Low Temp. Phys.* **122**, 591-604 (2001).

40. Das, D. *et al*. Magnetic field driven complex phase diagram of antiferromagnetic heavy-fermion superconductor Ce$_3$PtIn$_{11}$. *Sci. Rep.* **8**, 16703 (2018).

41. Watanabe, S. & Miyake, K. Roles of critical valence fluctuations in Ce- and Yb-based heavy fermion metals. *J. Phys.: Condens. Matter* **23**, 094217 (2001).

42. Yuan, H. Q. *et al*. Observation of two distinct superconducting phases in CeCu$_2$Si$_2$. *Science* **302**, 2104-2107 (2003).

43. Seyfarth, G. *et al*. Heavy fermion superconductor CeCu$_2$Si$_2$ under high pressure: Multiprobing the valence crossover. *Phys. Rev. B* **85**, 205105 (2012).

44. Wang, H. H. *et al*. Anomalous connection between antiferromagnetic and superconducting phases in the pressurized noncentrosymmetric heavy-fermion compound CeRhGe$_3$. *Phys. Rev. B* **99**, 024504 (2019).

## Acknowledgements

We thank G. M. Zhang and C. Petrovic for their helpful discussions. This work is supported by the Ministry of Science and Technology of China (Grants No. 2016YFA0300503, No. 2017YFA0303103), the NSAF (Grant No. U1630248), the National Natural Science Foundation of China (Grant No. 11674367, No. 11774401), the Shanghai Municipal Science and Technology Major Project (Grant No. 2019SHZDZX01), and the Zhejiang Provincial Natural Science Foundation (Grant No. LZ18A040002). DCP is supported by the Chinese Academy of Sciences through 2018PM0036.

## Author Contributions

S.Y.L. and E.J.C. conceived the idea and designed the experiments. E.J.C. was responsible for single-crystal growth and ambient/high-pressure transport experiments. C.C.Z., T.P.Y., J.M.N., B.L.P., Y.Y.H. and L.S.W. participated in the collection of data. T.P.Y. conducted the Rietveld refinement of powder X-ray diffraction. E.J.C., S.Y.L., Y.F.Y., Y.J.X., T.P.Y. and D.C.P. analyzed the data. E.J.C., D.C.P., S.Y.L. and Y.F.Y. wrote the manuscript. All authors discussed the results and commented on the manuscript.

## Competing interests

The authors declare no competing interests.

## Additional Information

**Supplementary information** is available for this paper at URL inserted when published.

**Correspondence** and requests for materials should be addressed to Y.-F.Y. (yifeng@iphy.ac.cn) or S.Y. L. (shiyan_li@fudan.edu.cn).

Figure 1

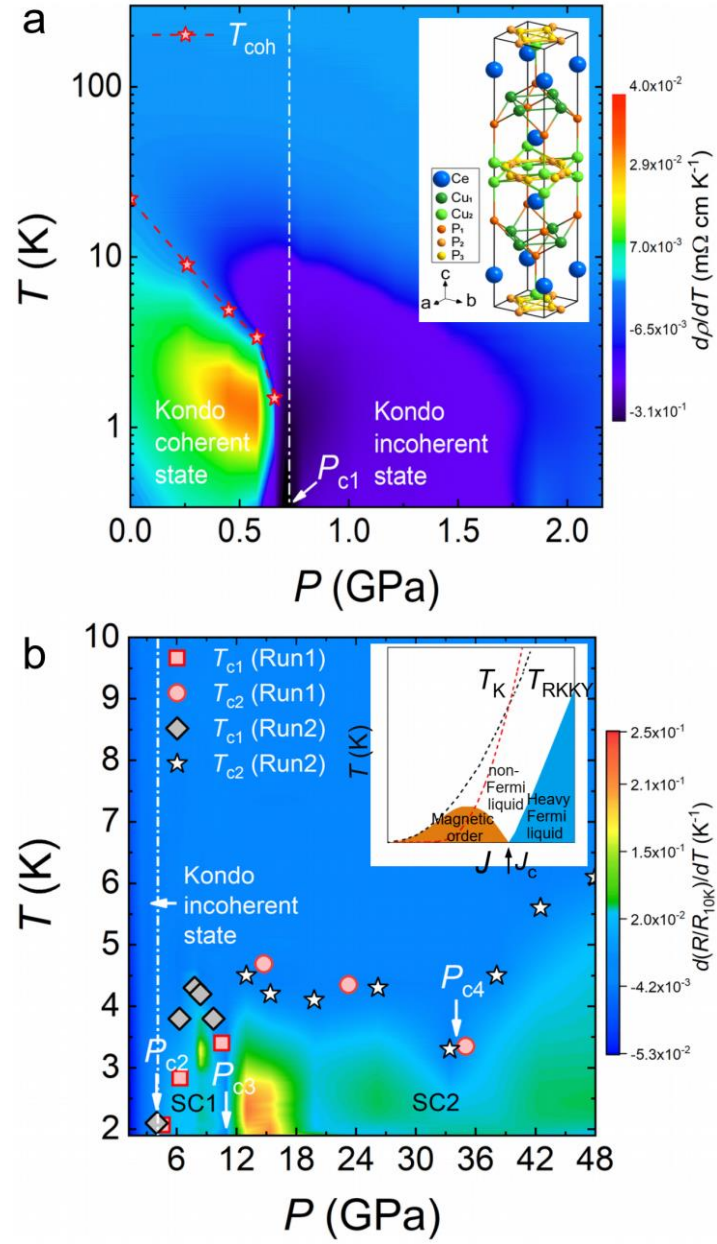

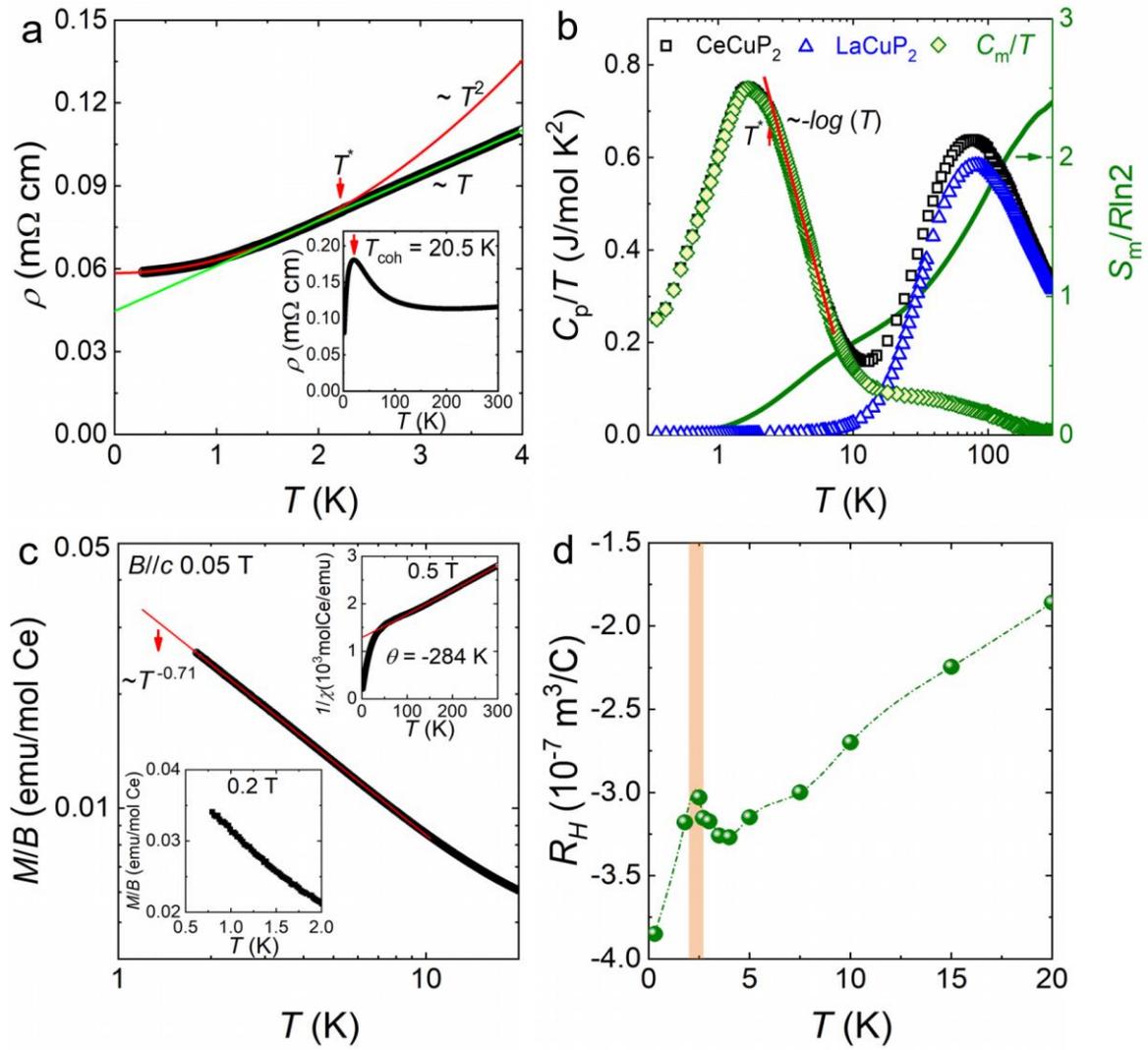

Figure 3

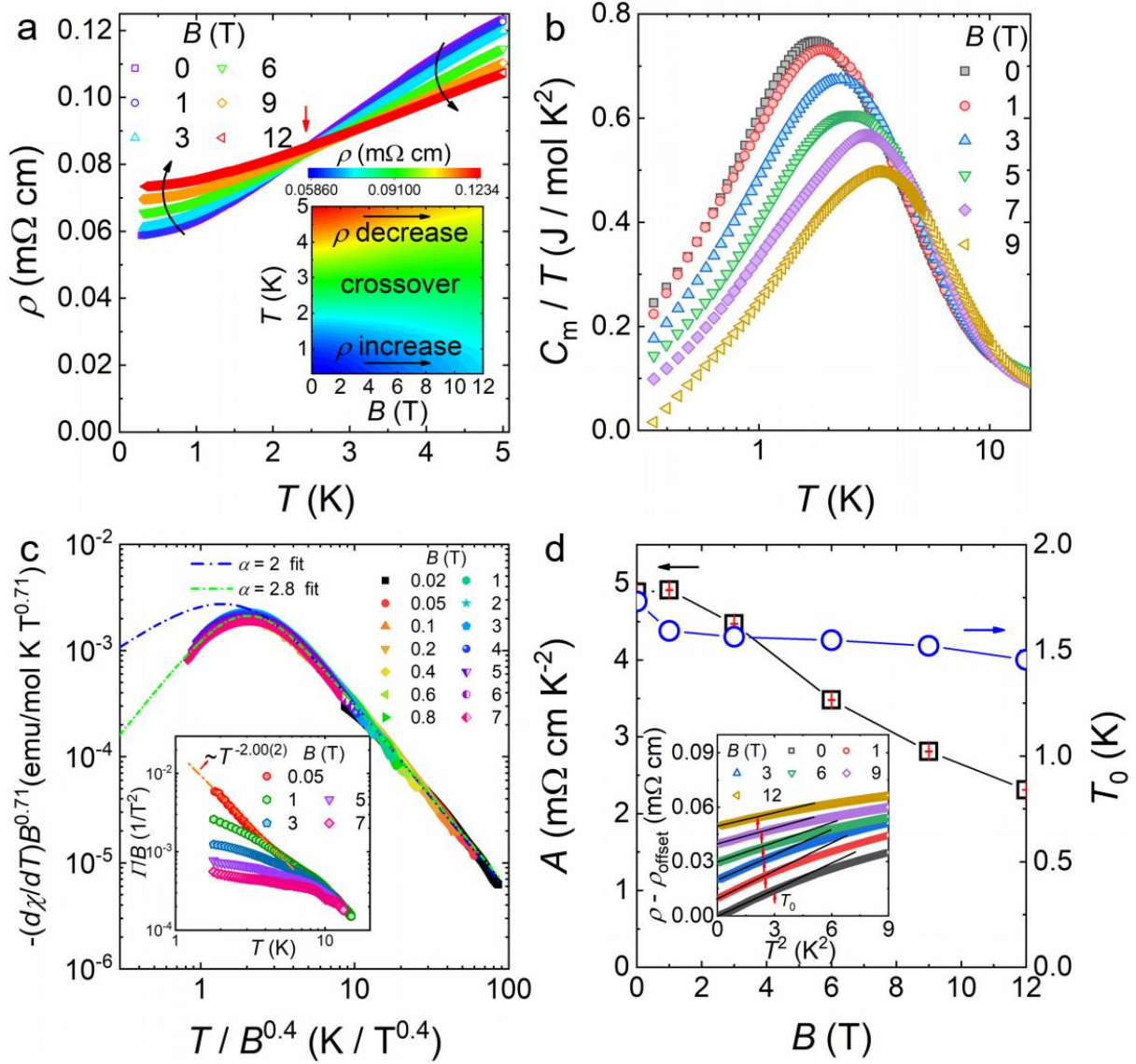

Figure 4

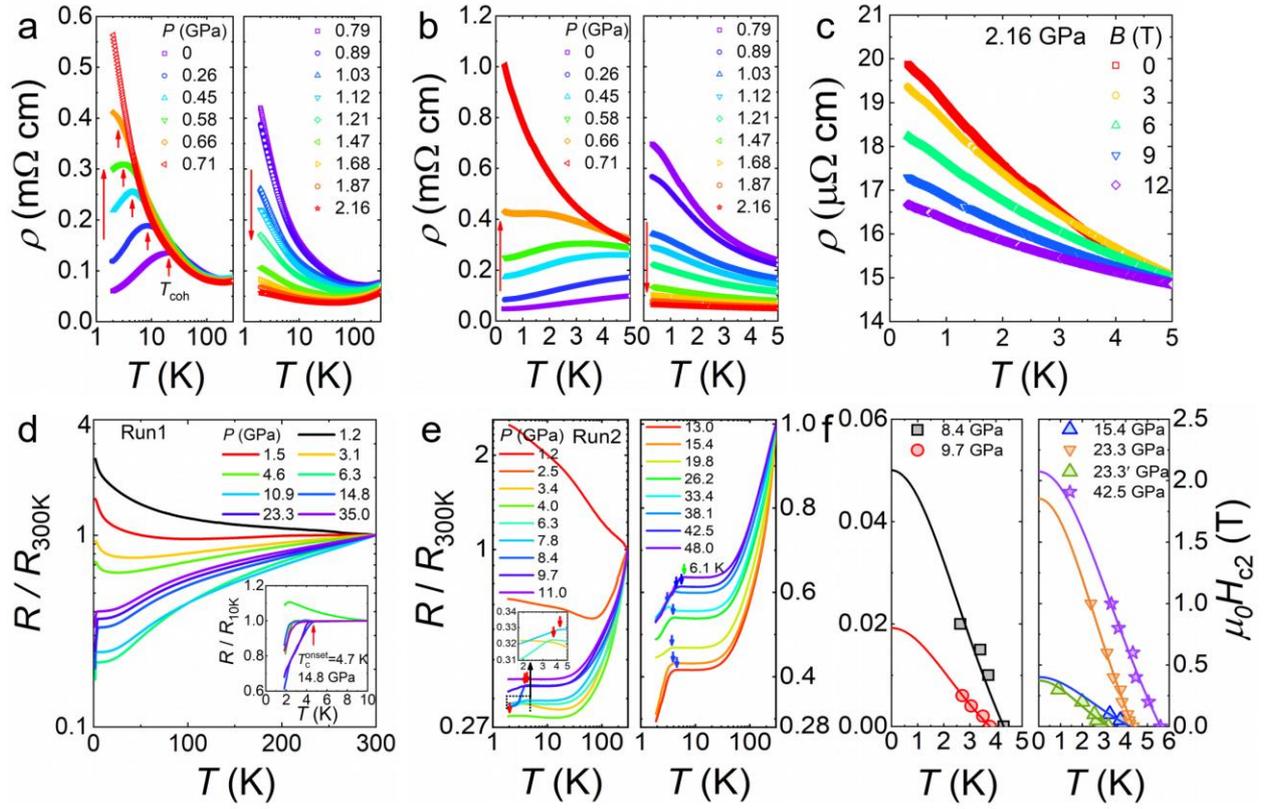

**Figure captions**

**Figure 1. Pressure-temperature (*P-T*) phase diagrams of CeCuP$_2$. a** *P-T* phase diagram of the CeCuP$_2$ single crystal under low pressure. The pressure dependence of coherence temperature ($T_{coh}$) taken from Fig. 4(a) is added, and the color represents the magnitude of $d\rho/dT$. For pressures $P < P_{c1}$, the system remains in a Kondo coherent state. For $P > P_{c1}$, a Kondo incoherent state takes over. Inset shows the crystal structure of CeCuP$_2$. CeCuP$_2$ crystallizes in a tetragonal structure (space group: *I*4/*mmm* (No. 139)) with three different P sites, two differing Cu sites but only one Ce atomic position. **b** *P-T* phase diagram of CeCuP$_2$ powder obtained by crushing single crystals, under higher pressure. Run1 and Run2 denote resistance measurements without pressure transmitting medium and with NaCl as the pressure transmitting medium, respectively. For 1.2 GPa $< P < P_{c2}$, the system remains in a Kondo incoherent state. At $P_{c2}$, superconductivity appears. For $P_{c2} \leq P < P_{c3}$, the system is in the first superconducting state (SC1). At $P_{c3}$ ~ 11.0 GPa, the first SC1 disappears. With increasing pressure beyond $P_{c3}$, the second superconducting state (SC2) arises. Upon further compression to a critical pressure ($P_{c4}$) of ~ 34.0 GPa, the $T_c^{onset}$ defined according to the onset of the transition for SC2 reaches a minimum, and then it goes up to 6.1 K at 48.0 GPa. The background color represents the magnitude of $d(R/R_{10K})/dT$ taken from Run2 measurements. Inset shows the regular Doniach phase diagram [7].

**Figure 2. Physical properties of CeCuP$_2$ single crystal under ambient pressure. a** In-plane resistivity of a CeCuP$_2$ single crystal from 0.3 to 4 K in zero field. The green and red lines represent linear and quadratic fits, respectively, and indicates that the ground state of CeCuP$_2$ is a Fermi liquid. A characteristic temperature ($T^*$) denotes the crossover from non-Fermi liquid to Fermi liquid

behavior. Inset shows the resistivity up to 300 K, which displays typical Kondo heavy-fermion behavior with a coherence temperature of 20.5 K. **b** Specific heats of CeCuP$_2$, LaCuP$_2$ and magnetic specific heat ($C_m/T$) extracted by deducting the electron and phonon contributions in LaCuP$_2$ from those in CeCuP$_2$. The red line represents the $-\log(T)$ dependence. The green line denotes the magnetic entropy ($S_m$) calculated based on $C_m/T$. The red arrow marks the characteristic temperature ($T^*$) where it deviates from $-\log(T)$, corresponding to that in the resistivity. **c** Temperature dependence of the *dc* susceptibility measured at 0.05 T. The solid line shows the low-field fitting with a $T^{-0.71}$ divergence. Right inset shows the reciprocal susceptibility from 1.8 to 300 K measured at 0.5 T, and the red solid line represents the Curie-Weiss fit, which yields a Weiss temperature ($\theta$) of -284 K, indicating antiferromagnetic interactions. Left inset shows the low-temperature data at 0.2 T below 2 K, excluding long-range magnetic order. **d** Hall coefficient ($R_H$) obtained by fitting the data of Hall resistivity from 0 to 3 T. The negative $R_H$ indicates that electron carriers dominate. On cooling from 20 to 4 K, the magnitude of $R_H$ increases monotonously. Subsequently, a peak situating ~2.4 K emerges abruptly. The shadowed area marks the change in the $R_H$. The carrier concentration at 0.3 K can be estimated to be $1.97 \times 10^{19}$ cm$^{-3}$.

**Figure 3. Physical properties of CeCuP$_2$ single crystal under magnetic field. a** In-plane resistivity of CeCuP$_2$ under magnetic fields applied along the *c* axis. Under magnetic fields, the resistivity curves meet in a point at ~ 2.4 K, below which the magnetoresistance (MR) is positive, and above which it is negative. Inset shows a contour plot of the temperature dependence of the resistivity at several magnetic fields; the green area represents the crossover where the MR changes sign. **b** Magnetic specific heat of CeCuP$_2$ single crystal under magnetic field applied along the *c* axis.

**c** Scaling collapse of the temperature derivative of $\chi$ multiplied by $B^{0.71}$ as a function of the variable $T/B^{0.4}$. Inset shows the Grüneisen ratio $\Gamma$ divided by $B$. **d** Field dependence of the $T^2$ coefficient $A$ (in $\Delta\rho = AT^2$), and the upper limit of the $T^2$ range, $T_0$ (arrows in the inset). Inset displays the temperature dependence of the resistivity at low temperature, plotted as $\rho(T)$-$\rho_{\text{offset}}$ vs. $T^2$, at several magnetic fields. $\rho_{\text{offset}}$ is a constant arbitrary offset chosen for clarity of display. The solid lines are linear fits to the data below $T_0$. The slope of these lines is the inelastic electron-electron scattering coefficient $A$.

**Figure 4. Electrical transport measurements of CeCuP$_2$ under pressure, and emergence of superconductivity.** Resistivity of a CeCuP$_2$ single crystal at different pressures, (**a**) from 2 to 300 K and (**b**) from 0.3 to 5 K. Upon increasing the pressure from 0 to 0.71 GPa, the Kondo coherence temperature monotonically decreases. Above 0.71 GPa, the resistivity decreases with increasing pressure. **c** Temperature dependence of the resistivity under 2.16 GPa at several magnetic fields. With increasing field, the resistivity monotonically decreases, in contrast to the behavior at ambient pressure, as plotted in Fig. 3(a). **d** Temperature dependence of the normalized resistance of CeCuP$_2$ powder obtained by crushing single crystals obtained by crushing a single crystal at higher pressures without pressure transmitting medium (denoted as Run1). Inset shows the superconducting transitions, and the arrow marks the onset transition temperature $T_c^{\text{onset}}$ of 4.7 K at 14.8 GPa. **e** Resistance measurements of CeCuP$_2$ powder obtained by crushing single crystals at higher pressures with NaCl as the pressure transmitting medium (denoted as Run2). **f** Temperature dependence of the upper critical field $\mu_0 H_{c2}$ at several pressures. 23.3ˊGPa in right panel denotes the data that superconducting transition temperature is defined according to a 10% drop ($T_c^{10\%}$). Lines represent the Ginzburg-Landau fit.

# Supplementary Information to

# "Quantum phase transitions and superconductivity in the pressurized heavy-fermion compound CeCuP$_2$"

Erjian Cheng[1], Chuchu Zhu[1], Tianping Ying[1], Yuanji Xu[2], Darren C. Peets[3], Jiamin Ni[1], Binglin Pan[1], Yeyu Huang[1], Linshu Wang[1], Yi-feng Yang[2,*] and Shiyan Li[1,4,5,*]

[1] *State Key Laboratory of Surface Physics, Department of Physics, Fudan University, Shanghai 200438, China*

[2] *Beijing National Laboratory for Condensed Matter Physics and Institute of Physics, Chinese Academy of Sciences, Beijing 100190, China*

[3] *Ningbo Institute of Materials Technology and Engineering, Chinese Academy of Sciences, Ningbo, Zhejiang 315201, China*

[4] *Collaborative Innovation Center of Advanced Microstructures, Nanjing 210093, China*

[5] *Shanghai Research Center for Quantum Sciences, Shanghai 201315, China*

## Supplementary Note 1: Chemical composition and crystal structure determinations

The chemical composition of the CeCuP$_2$ single crystals is Ce : Cu : P = 1 : 1.14 : 2 as determined by inductively coupled plasma (ICP) spectroscopy. We performed Rietveld refinement of powder X-ray diffraction (XRD) data obtained by crushing several single crystals of CeCuP$_2$, as shown in Supplementary Fig. 1(a). Fractional occupancies are expected for the Cu2, P2, and P3 atoms [1]. For Rietveld refinement, we use the results of ICP, i.e., Ce : Cu = 1 : 1.14, as fixed parameters, while the P stoichiometry is taken as a free parameter because of its higher uncertainty as a light element. The Rietveld refinement of CeCuP$_2$ gives the refined crystallographic parameters shown in Supplementary Table 1, with $R_p$ = 4.17, $R_{wp}$ = 6.46, $R_{exp}$ = 4.01 and $\chi^2$ = 2.6. Based on the refinement, the P stoichiometry is 2.186(6), yielding Ce : Cu : P = 1 : 1.14 : 2.186(6), which is slightly higher than ICP. The crystal structure shown in the main text (inset to Fig. 1(a)) is based on this refinement. The crystal structure of LaCuP$_2$ has also been checked by XRD, and the chemical composition of LaCuP$_2$ was determined to be 1 : 1.11 : 1.9 by EPMA.

The largest surface of a CeCuP$_2$ single crystal is indexed to the (00$l$) plane, as plotted in Supplementary Fig. 1(b). The inset to Supplementary Fig. 1(b) shows the X-ray rocking curve of the (008) peak with the full width at half maximum (FWHM) of 0.08 °, indicative of the high quality of our samples.

| Compound | | CeCuP$_2$ (space group: $I4/mmm$ (139)) | |
|---|---|---|---|
| Atoms | (x/a, y/b, z/c) | occupancy ratio | Wyckoff Positions |
| Ce | (0, 0, 0.11845(5)) | 1.000(0) | 4e |
| Cu1 | (0.5, 0, 0.25) | 1.000(0) | 4d |
| Cu2 | (0, 0, 0.46424(8)) | 0.140(0) | 4e |
| P1 | (0, 0, 0.32038(3)) | 1.000(0) | 4e |
| P2 | (0.5, 0, 0) | 0.654(2) | 4c |
| P3 | (0.31867(1), 0, 0.5) | 0.266(2) | 8j |
| Cell parameters | | Refinement parameters | |
| $a$ (Å) | 3.90663(6) | $R_p$ | 4.17 |
| $b$ (Å) | 3.90663(6) | $R_{wp}$ | 6.46 |
| $c$ (Å) | 19.6578(4) | $R_{exp}$ | 4.01 |
| $V$ (Å$^3$) | 300.012(8) | $\chi^2$ | 2.6 |

**Supplementary Table 1. Crystallographic parameters of CeCuP$_2$.**

## Supplementary Note 2: XRD measurements

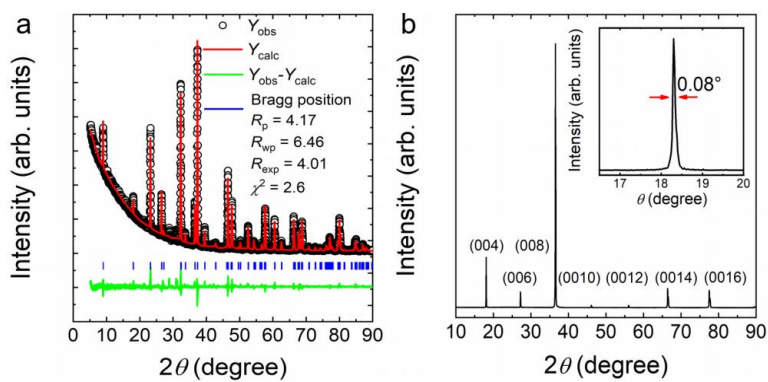

**Supplementary Figure 1. XRD measurements of CeCuP$_2$. a** Rietveld refinement of powder XRD of CeCuP$_2$ obtained by crushing several single crystals. The Rietveld refinement successfully converges to $R_p$ = 4.17, $R_{wp}$ = 6.46, $R_{exp}$ = 4.01 and 2.6. **b** XRD pattern from the largest surface of a CeCuP$_2$ single crystal. Only (00$l$) Bragg peaks are observed. Inset shows the X-ray rocking curve of the (008) peak.

## Supplementary Note 3: Resistivity of LaCuP$_2$

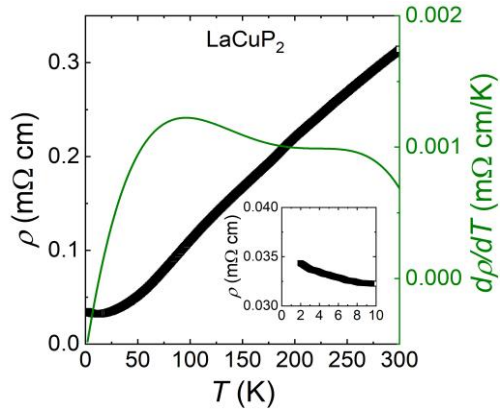

**Supplementary Figure 2. Resistivity of polycrystalline LaCuP$_2$ at ambient pressure.** The resistivity shows a metallic behavior from 10 to 300 K, and then displays semi-metallic behavior below 10 K. The derivative result of resistivity has also been displayed, and one can see an anomaly around 100 K, which may be related to the charge-density-wave (CDW) transition, but more evidence would be required. To check this, Hall resistivity and detailed theoretical analysis are called for.

**Supplementary Note 4: Specific heat measurements under magnetic field**

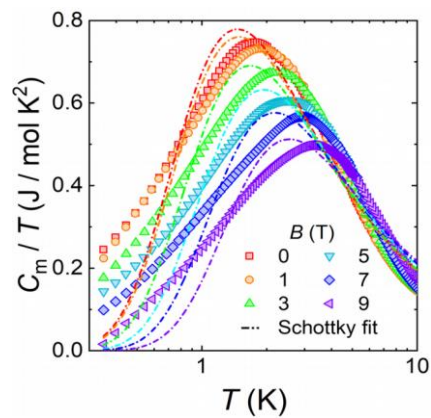

**Supplementary Figure 3. The $C_m/T$ of a CeCuP$_2$ single crystal at various magnetic fields.** The broken lines represent the fits according to a two-level Schottky function [3]. Obviously, the hump in CeCuP$_2$ contradicts the prediction of the Schottky

function for all fixed magnetic fields.

## Supplementary Note 5: *dc* and *ac* magnetic susceptibility measurements

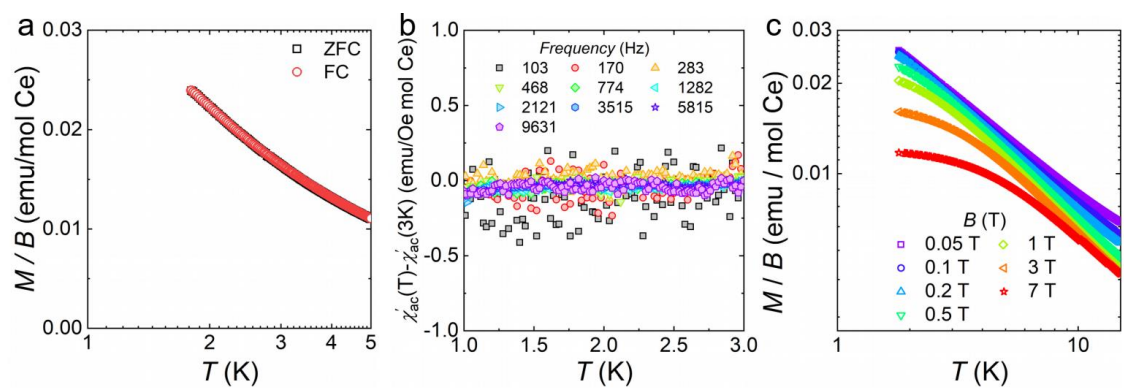

**Supplementary Figure 4.** *dc* **and** *ac* **magnetic susceptibility of CeCuP$_2$ single crystal. a** Zero field cooling (ZFC) and field cooling (FC) of *dc* magnetic susceptibility of CeCuP$_2$ single crystal from 1.8 K to 5 K with magnetic field applied along *c* axis. **b** *ac* magnetic susceptibility of CeCuP$_2$ single crystal. From *dc* and *ac* magnetic susceptibility, long- and short-range magnetic transitions around 1.8 K can be excluded. **c** Temperature dependence of the *dc* susceptibility under several magnetic fields.

## Supplementary Note 6: Hall measurements

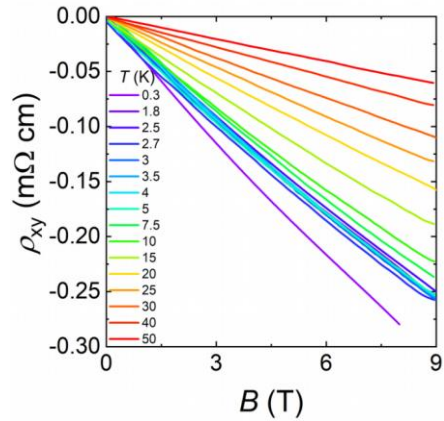

**Supplementary Figure 5. Hall measurements of a CeCuP$_2$ single crystal at several fixed temperatures.**

## Supplementary Note 7: Magnetoresistance under ambient and high pressure

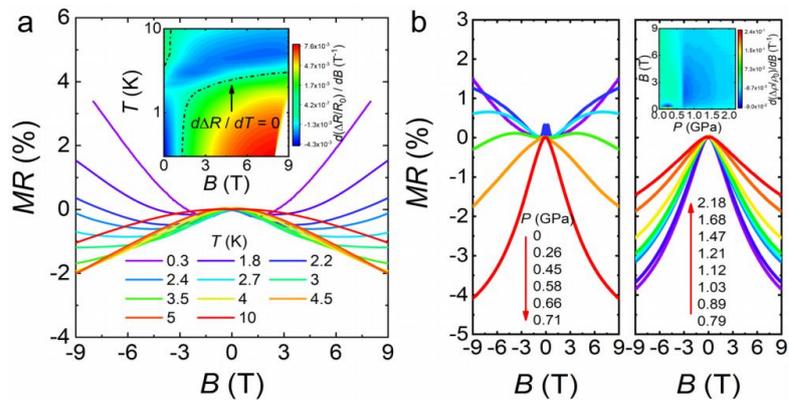

**Supplementary Figure 6. Magnetoresistance (MR) of a CeCuP$_2$ single crystal as a function of temperature. a** Under ambient pressure and **b** high pressures with magnetic field applied along *c* axis.

To clarify the behavior of magnetic Ce 4f electrons, the pressure dependence of the magnetoresistance (MR) has been investigated at 2 K, as displayed in Supplementary Fig. 6(b). With increasing pressure, the MR changes from positive to negative gradually, and the curvature of the MR systematically changes. At 0.71 GPa and 9 T, the magnitude of the MR reaches a maximum. The changes in the MR from 0 to 0.71 GPa are associated with the evolution of $T_{coh}$. At higher pressures, the MR remains negative, but its magnitude decreases. The inset in the right panel of Supplementary Fig. 6(b) shows a contour plot of the *P-B* phase diagram, and there is a clear boundary at ∼ 0.7 GPa, indicative of the change of the scattering mechanism under pressure, possibly associated with magnetic fluctuations. It is also worth noting that the MR in low magnetic field shows a pronounced cusp at 0.26 GPa, suggestive of a possible weak anti-localization (WAL) effect, but to verify this, Hall resistivity and detailed theoretical analysis are called for.

We attribute the change of MR from positive to negative under ambient pressure to a nonmagnetic crossover rather than magnetic fluctuations. Because, in heavy-fermion materials, the competition between a decreasing *T*-dependent term and an increasing *T*-independent residual resistivity contribution will yield a maximum in the MR of a Kondo lattice [4,5]. At ambient pressure, the MR for $CeCuP_2$ shows a monotonic field dependence, rather than a maximum for all fixed temperatures (see Supplementary Fig. 6(a)), whereas, under pressure, the maximum in MR can be found at several lower pressures, characteristic of the heavy-fermion behavior, i.e., the Ce 4f electrons. Additionally, the temperature-dependent MR at 2.16 GPa (Fig. 4(c)) is in striking contrast to that at ambient pressure (Fig. 3(a)), also implying the different origin of scattering mechanism.

## Supplementary Note 8: Pressure dependence of the electron-electron scattering coefficient $A$

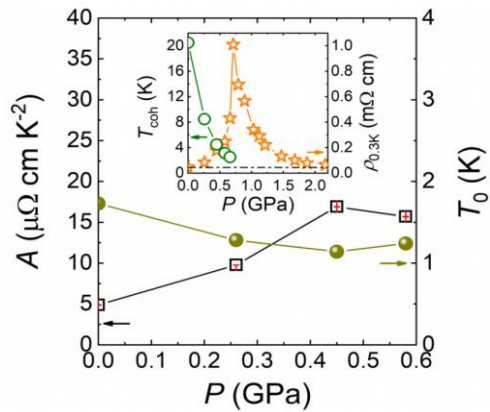

**Supplementary Figure 7. Pressure dependence of the $T^2$ coefficient $A$, and the upper limit of the $T^2$ range, $T_0$.** Inset shows the pressure dependence of the $T_{coh}$ and the resistivity at 0.3 K.

## Supplementary Note 9: Magnetic field dependence of the superconducting transition at several selected pressures

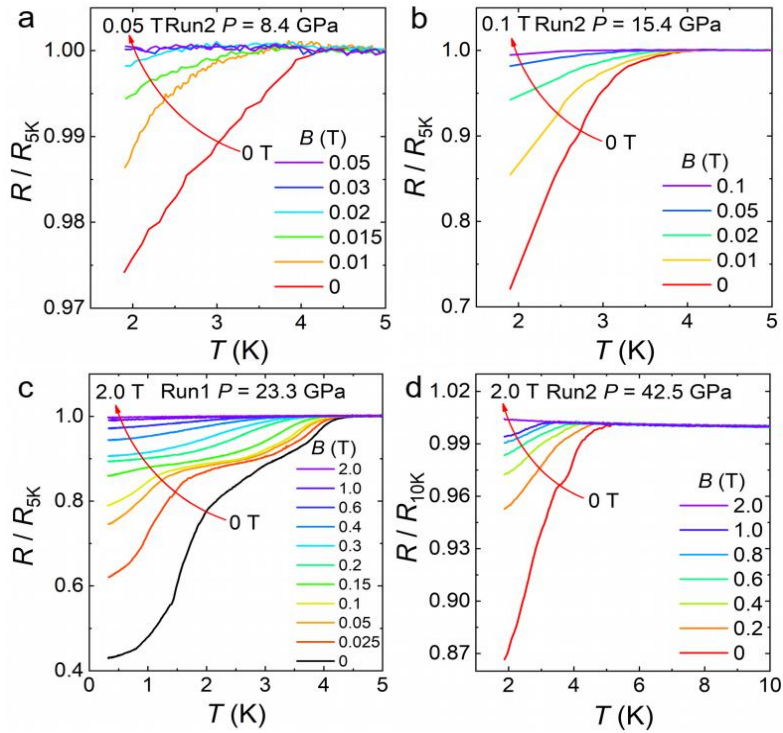

**Supplementary Figure 8. Magnetic field dependence of the superconducting transition at several selected pressures for CeCuP$_2$ powder obtained by crushing single crystals**—**a** 8.4 GPa; **b** 15.4 GPa; **c** 23.3 GPa; **d** 42.5 GPa. Increasing the magnetic field gradually suppresses the superconducting transition.

# Supplementary Note 10: Comparison of the upper critical field and the Pauli field for Ce-based heavy-fermion superconductors

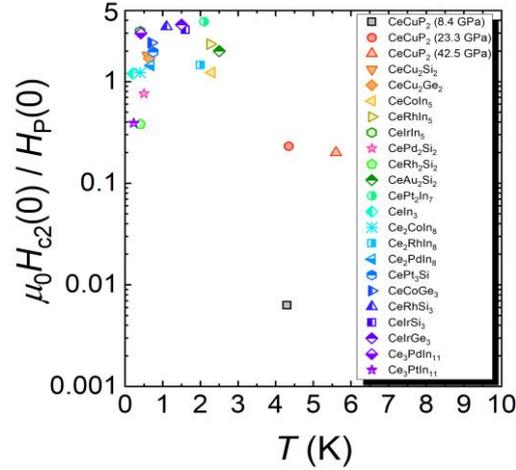

**Supplementary Figure 9. Comparison of the upper critical field ($\mu_0 H_{c2}(0)$) and the Pauli field ($H_p(0)$) for Ce-based heavy-fermion superconductors.** Data is taken from Refs. 6, 7 and 8 for other Ce-based heavy-fermion superconductors.

# Supplementary Note 11: Scaling analysis of low-temperature resistivity at pressures near $P_{c1}$

To argue the pressure-induced valence change in $CeCuP_2$, a scaling analysis of resistivity has been conducted according to the scaling method proposed in $CeCu_2Si_2$ [9] and developed in $CeRhGe_3$ [10]. Different from $CeCu_2Si_2$ and $CeRhGe_3$, the resistivity above $P_{c1}$ in $CeCuP_2$ is divergent on cooling, which is difficult to extract the impurity scattering contribution (i.e., residual resistivity $\rho_0$). Compared with the scattering of spin singlets in the incoherent state in $CeCuP_2$, the impurity scattering contribution can be ignored. Supplementary Fig. 9(a) plots the pressure-dependent resistivity at selected temperatures. Consistent with $CeCu_2Si_2$ [9] and $CeRhGe_3$ [10], the $\rho$ begins to drop significantly above $P_{c1}$, implying an increased delocalization of the 4$f$ electrons in CeCuP$_2$. To better show the delocalization effect, a normalized resistivity $\rho_{nor}$ ($\rho_{nor}$ = [$\rho(P)$ - $\rho(P_{c1})$]/$\rho(P_{c1})$) is defined, and the pressure-dependent $\rho_{nor}$ is plotted in Supplementary Fig. 10(b). A pressure $P_{vc}$ is also defined, at which $\rho_{nor}$ drops by 50 % from its value at $P_{c1}$, as shown in the inset of Fig. S10(b). By extrapolating the $P_{vc}$ data to zero temperature (inset to Supplementary Fig. 10(c)), a critical pressure $P_{cr}$ ~1 GPa can be obtained. Supplementary Figure 10(c) shows the slope $\chi$ at $P_{vc}$. We fit the low-temperature points through the form $\chi \propto (T-T_{cr})^{-1}$, and obtain a critical temperature $T_{cr}$ of -1.1(2) K. We also introduce a generalized distance $h/\theta$, where $h = (P-P_{vc})/P_{vc}$, and $\theta = (T - T_{cr})/|T_{cr}|$, and the $\rho_{nor}$ as a function of $h/\theta$ is plotted in Fig. S10(d). As in CeCu$_2$Si$_2$ [9] and CeRhGe$_3$ [10], the data below 10 K in CeCuP$_2$ shows a $h/\theta$ collapse, evidencing that the pressurized CeCuP$_2$ at $P_{c1}$ is in proximity to a critical valence instability.

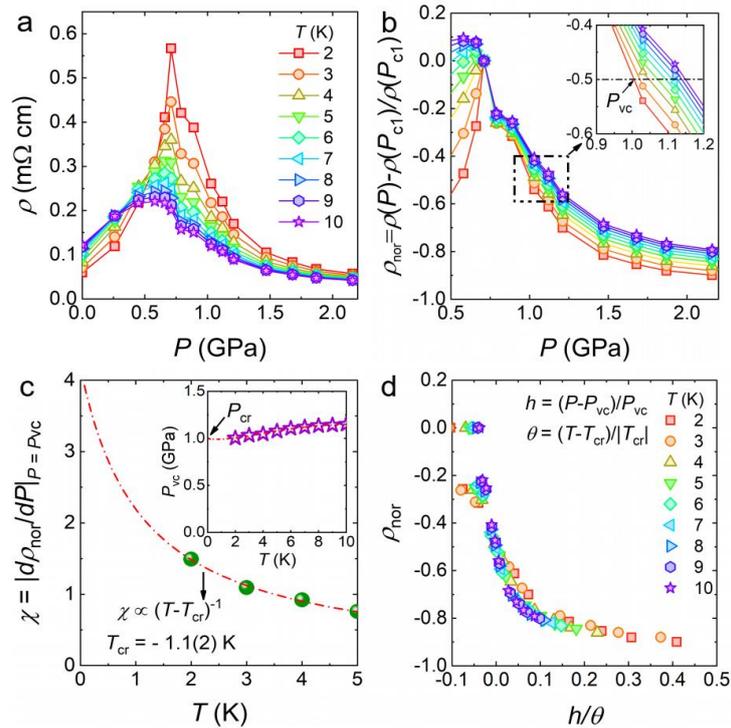

**Supplementary Figure 10. Scaling analysis of low-temperature resistivity of a CeCuP$_2$ single crystal at pressures near $P_{c1}$ ~0.7 GPa. a** Pressure-dependent resistivity at selected temperatures. **b** Normalized resistivity $\rho_{nor}$ as a function of pressure, where $\rho_{nor} = [\rho(P) - \rho(P_{c1})]/\rho(P_{c1})$. Inset shows the definition of the pressure $P_{vc}$, at which $\rho_{nor}$ drops by 50 % from its value at $P_{c1}$. **c** Temperature-dependent the slope $\chi$ at $P_{vc}$ in $\rho_{nor}$-$P$ curves. The red dashed line denotes a fit of $\chi \propto (T-T_{cr})^{-1}$, yielding a critical temperature $T_{cr}$ of -1.1(2) K. Inset displays the temperature-dependent $P_{vc}$. By extrapolating to zero temperature, a critical pressure $P_{cr}$ of ~ 1 GPa can be obtained. **d** Collapse of $\rho_{nor}$ as a function of generalized distance $h/\theta$, where $h = (P-P_{vc})/P_{vc}$, and $\theta = (T - T_{cr})/|T_{cr}|$.

# Supplementary Note 12: The temperature dependence of the resistivity at several pressures

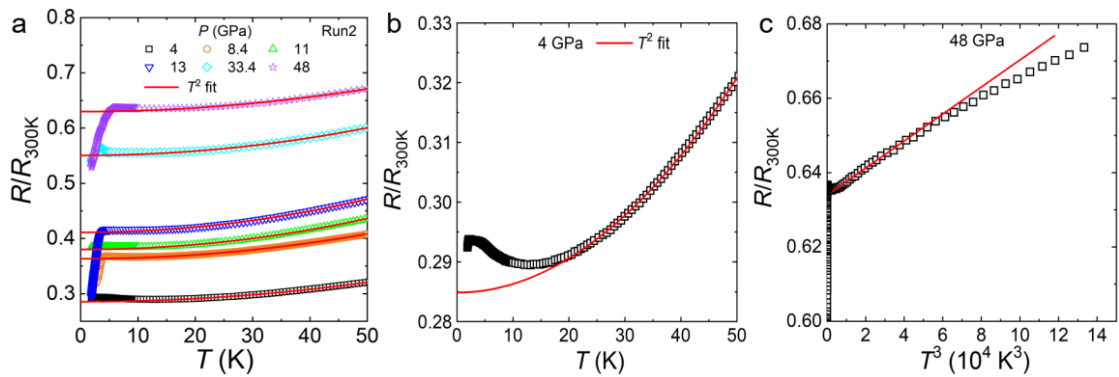

**Supplementary Figure 11. Fit to the temperature dependence of the resistivity ($R/R_{300K} = \rho/\rho_{300K}$) at several pressures in Run2 measurements.** Data are taken from Fig. 4(e). **a** The low-temperature resistivity ranging from ~ 15 K to 50 K

displays nearly $T^2$ dependence. **b** The resistivity at 4 GPa shows a low-temperature upturn, where superconductivity arises from. **c** The lower-temperature resistivity at 48 GPa displays $T^3$ dependence. The temperature dependence of the resistivity is beyond the self-consistent renormalization (SCR) picture.

## Supplementary References

1. Chykhrij, S. I., Loukashouk, G. V., Oryshchyn, S. V., Kuz'ma, Y. B. Phase equilibria and crystal structure of compounds in the Ce-Cu-P system, *J. Alloy. Compd.* **248**, 224-232 (1997).

2. Mao, H. -K., Xu, J. A. & Bell, P. Calibration of the ruby pressure gauge to 800-kbar under quasi-hydrostatic conditions. *J. Geophys. Res.* **91**, 4673-4676 (1986).

3. Moler, K. A. Specific heat of YBa$_2$Cu$_3$O$_{7-\delta}$. *Phys. Rev. B* **55**, 6 (1996).

4. Ohkawa, F. J. Magnetoresistance of Kondo lattices. *Phys. Rev. Lett.* **64**, 2300 (1989).

5. Hu, T. *et al*. Non-Fermi liquid regimes with and without quantum criticality in Ce$_{1-x}$Yb$_x$CoIn$_5$. *P. Natl. Acad. Sci. USA* **110**, 7160 (2013).

6. Weng, Z. F. *et al*. Multiple quantum phase transitions and superconductivity in Ce-based heavy fermions. *Rep. Prog. Phys.* **79**, 094503 (2016).

7. Das, D. *et al*. Magnetic field driven complex phase diagram of antiferromagnetic heavy-fermion superconductor Ce$_3$PtIn$_{11}$. *Sci. Rep.* **8**, 16703 (2018).

8. Kratochvíová, M. *et al*. Coexistence of antiferromagnetism and superconductivity in heavy fermion cerium compound Ce$_3$PdIn$_{11}$. *Sci. Rep.* **5**, 15904 (2015).

9. Seyfarth, G. *et al*. Heavy fermion superconductor $CeCu_2Si_2$ under high pressure: Multiprobing the valence crossover. *Phys. Rev. B* **85**, 205105 (2012).

10. Wang, H. H. *et al*. Anomalous connection between antiferromagnetic and superconducting phases in the pressurized noncentrosymmetric heavy-fermion compound $CeRhGe_3$. *Phys. Rev. B* **99**, 024504 (2019).